\documentstyle[sprocl,epsfig]{article}

\def\be{\begin{equation}}
\def\ee{\end{equation}}
\def\bea{\begin{eqnarray}}
\def\eea{\end{eqnarray}}

\begin{document}
\begin{flushleft}
{\normalsize \tt
SFB/CPP-04-39
\\
DESY 04-160
\\ September 2004}
\end{flushleft}
\vspace*{.2cm}

\title{ON MASTER INTEGRALS \\ FOR TWO LOOP BHABHA SCATTERING
 \footnote{Work supported in part 
  by European's 5-th Framework under contract HPRN--CT--2000--00149 Physics at
  Colliders, 
  by TMR, EC-Contract No. HPRN-CT-2002-00311 (EURIDICE), 
  by Deutsche Forschungsgemeinschaft under contract SFB/TR 9--03, 
  and by the Polish State Committee for Scientific Research (KBN)
  for the research project in years 2004-2005.}\\}

\author{M. CZAKON$^{1,2}$, J. GLUZA$^{1,2}$, T. RIEMANN$^1$\\}
\vspace{.4cm}
\address{$^1$Deutsches Elektronen-Synchrotron, DESY,
   Platanenallee 6, 15738 Zeuthen, Germany\\
$^2$ Institute of Physics, University of Silesia, \\
   ul. Uniwersytecka 4, 40007 Katowice, Poland}


\maketitle\abstracts{
All scalar master integrals (MIs) for 
massive 2-loop QED Bhabha scattering are identified. 
The 2- and 3-point
MIs have been calculated in terms of harmonic polylogarithms 
with the differential equation method. 
The calculation of 4-point MIs is underway. 
We sketch some alternative methods which help to solve (mainly) 
singularities of some MIs.
}
\section{Introduction}
Bhabha scattering, $e^+e^- \to e^+e^-(n \gamma)$,
is used in many accelerators for the determination of luminosity.
The NLO  corrections in the electroweak model are 
known\cite{ew}.
The lack of knowledge of the complete NNLO corrections in massive QED
has been one of the major sources of the theoretical error at
LEP\cite{jadach}, 
the other two are light fermion initial state pair production and the hadronic 
vacuum polarization.
Several calculations have been already done to get massive NNLO QED results for Bhabha 
scattering\cite{Bonciani:2003te,Bonciani:2003cj,Smirnov:2001cm,ll04}.
The first step in calculations is to identify MIs. For genuine vertices 
the task has been completed in\cite{Bonciani:2003te}.
The additional 3-point MIs coming from boxes and, more involved, the 4-point MIs themselves
are presented for the first time at this conference in the first part of the
talk. 
The second major step includes the evaluation of MIs. 
This process is underway.
Two  powerful methods are in use. 
One is based on a bottom-up approach using differential 
equations\cite{Kotikov:1991hm}. 
The other uses 
Mellin-Barnes representations for the calculation of each MI separately. 
The (semi)-analytical results are already known for some of the more complicated cases\cite{Smirnov:2001cm}.  
Further, there is a general numerical method to calculate MIs at
fixed kinematical Euclidean points\cite{Binoth:2000ps}.

For our complete set of MIs and the current 
status of calculations, see\cite{ll04} and the web page\cite{web-masters:2004nn}. 
Here we focus on a presentation of the second part 
of the talk, namely  some alternative cross-checks of analytical results. 
\section{Some cross-checks of analytical results}
\subsection{Algebraic relations between IR-divergent MIs}
\begin{figure}[hb] 
\epsfig{file=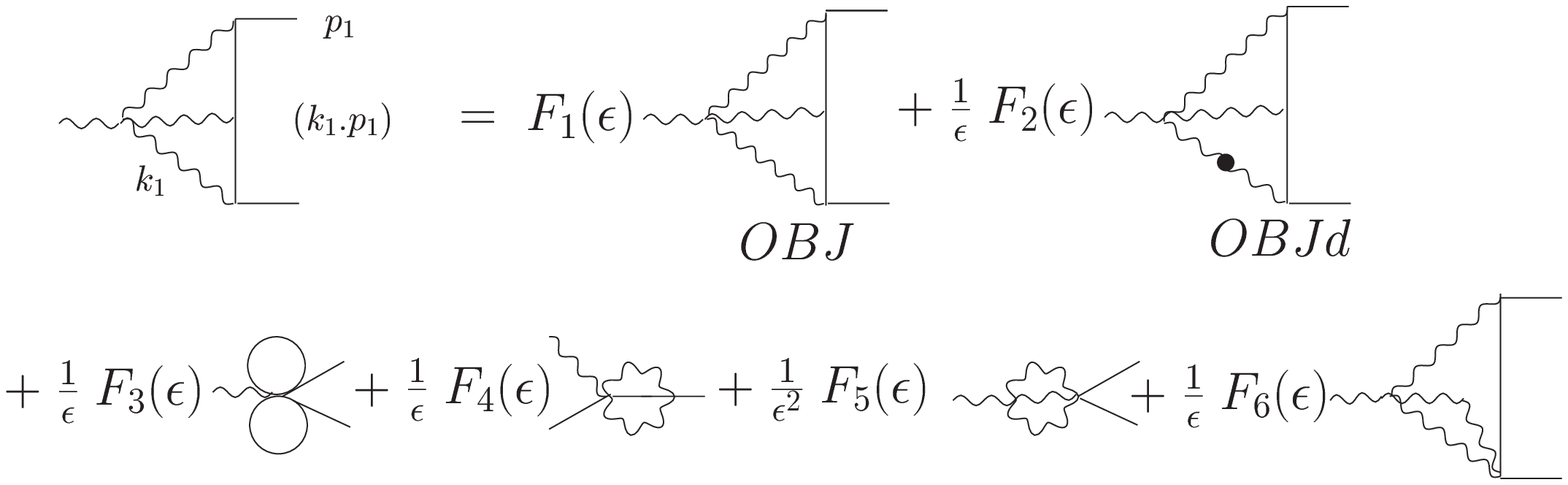,width=10cm,height=4cm}
     \caption{%
Relation between 
a vertex ${\it OBJ}$ (here V5l2m2 in terminology of$^{6,7}$), the same vertex with 
a dotted line ${\it OBJd}$ (here V5l2m2d), and 
with  irreducible numerator ($k_1p_1$).
%
} 
     \label{c1}
     \end{figure}

The method of determining MIs, which is realized in the package
IdSolver\cite{czak}, is based on 
the Laporta-Remiddi (LR) algorithm\cite{Laporta:1996mq}:
it determines and solves an appropriate set of algebraic equations with integration by parts\cite{Chetyrkin:1981qh}
and Lorentz invariance\cite{Gehrmann:1999as} identities. 
The result is a file, sometimes huge in size,  
which involves relations among 
MIs with different powers of propagators and 
irreducible numerators.  
Such relations can be used to fix singularities of purely IR divergent MIs.
An example is shown in Fig. \ref{c1}. 

\noindent
The diagrams  ${\it OBJ}$ and ${\it OBJd}$ are UV finite, but IR divergent. 
The corresponding diagram with an irreducible numerator is finite. 
If we expand the known MIs (second row in  Fig. \ref{c1}) into
series in 
$\epsilon = (4-d)/2$ with known coefficients $F_i/{\epsilon^{n_i}}$,
and make an ansatz for the unknowns, 
\begin{eqnarray}  
{\it OBJd}(s) &=& B_{-2}(s)\frac{1}{\epsilon^2}+B_{-1}(s)\frac{1}{\epsilon}+B_{0}(s)+\cdots ,
\end{eqnarray}
and similarly for ${\it OBJ}$, we get from Fig. \ref{c1} relations among the coefficients.
Here, e.g. ($a,b,c,d$ are known singularities of the appropriate simpler MIs):
\begin{eqnarray}  
0 = \frac{1}{\epsilon^3} \left[ a F_3(\epsilon)+b F_4(\epsilon)+c
  F_5(\epsilon)+d F_6(\epsilon)+F_2(\epsilon) B_{-2}(s) \right],
\end{eqnarray}
which allows to determine $B_{-2}(s) = {1}/{(4s)}$, in agreement with\cite{ll04}.
The crucial point is that there is an additional 
factor $1/\epsilon$ in front of ${\it OBJd}$ in comparison to ${\it OBJ}$.
To get e.g. $B_{-1}(s)$ or $A_{-2}(s)$ (analogous coefficient of ${\it OBJ}$),
another equation with a different, independent numerator  
would have to be used in addition.
\subsection{Exact subloop integration}
\begin{figure}[ht]
\epsfig{file=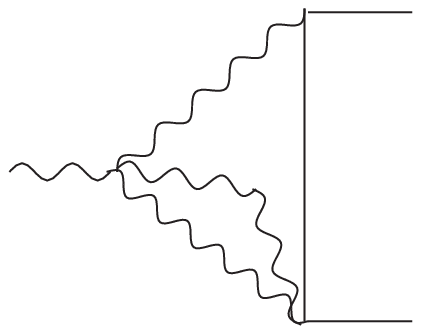,width=3.5cm,height=2cm}
     \caption{A diagram with a massless propagator subloop.}
     \label{c2}
     \end{figure}

The diagram in Fig. \ref{c2} has a massless UV divergent subloop. 


\hspace{3.4cm}
\begin{minipage}{7.8cm}
\vspace{-5.6cm}
$$=I=\int
\frac{d^dk_1d^dk_2}{[k_2^2][(k_1+k_2-p_1)^2][k_1^2+m^2][(k_1+p2)^2]}$$
\end{minipage}

\noindent
This allows to perform the subloop integration (over $k_2$) analytically,
resulting in a $C_0$-like
function  where 
one of the denominators appears with power $\epsilon$. 
Using a Feynman parameter representation, after integration over $k_1$
 and one of the Feynman parameters, 
we get ($t=(p_1+p_2)^2$):
\begin{eqnarray*}
I =
\pi^{d} \frac{\Gamma(1-\epsilon)^2\Gamma(\epsilon) }{16~\Gamma(1+\epsilon)^2 \Gamma(2-2\epsilon)}
\epsilon(1+\epsilon)
\frac{\Gamma\left(2\epsilon\right)}{\Gamma(2+\epsilon)}
I_{num}.
\end{eqnarray*}
This generates a singularity  in  $I_{num}$  at $x=0$, which 
can be treated by the following
subtraction ($F(x,y)=[(1-x)(1-y)^2+xy\frac{t}{m^2}]^{-2\epsilon}$):
\begin{eqnarray*}
I_{num} &=&
\int_0^1
\frac{dxdy~x^{-1+\epsilon}(1-x)^{1-2\epsilon}}{([(1-x)(1-y)^2m^2 +xyt]^2)^{2\epsilon}}\\
&=&
\frac{1}{(m^2)^{2\epsilon}} 
\left[
\frac{\Gamma(\epsilon)\Gamma(2-2\epsilon)}{\Gamma(2-\epsilon)}
\frac{1}{1-4\epsilon} + I_{reg}
\right],
\\
 I_{reg} &=& 
 \int_0^1 dx~x^{-1+\epsilon}(1-x)^{1-2\epsilon}\int_0^1 dy 
\left[ F(x,y)-F(0,y)\right].
\end{eqnarray*}
The remaining  integrations in $I_{reg} $ can be performed analytically or numerically after
the $\epsilon$-expansion:
\begin{eqnarray*}
 I_{reg} &=& 
\int_0^1 dx (1-x) e^{\epsilon \ln x}  e^{-2\epsilon \ln (1-x)}
\int_0^1 dy \frac{[\ln f(x,y)-\ln f(0,y)]}{x} 
\\ && \times~
\sum_{n=0}^{\infty} 
(-2\epsilon)^n 
\left[ \sum_{k=0}^n \ln^{n-k-1} f(x,y) \ln^{k} f(x,0) \right] = I_1 \epsilon+I_2 \epsilon^2+\cdots
\end{eqnarray*}
In this way both the singularities and also regular terms can be
obtained.
We have checked that they
coincide with our results in\cite{ll04}
and\cite{web-masters:2004nn} for the MI V4l1m2.
\subsection{Subtracting a counter term} 
\begin{figure}[ht] 
\epsfig{file=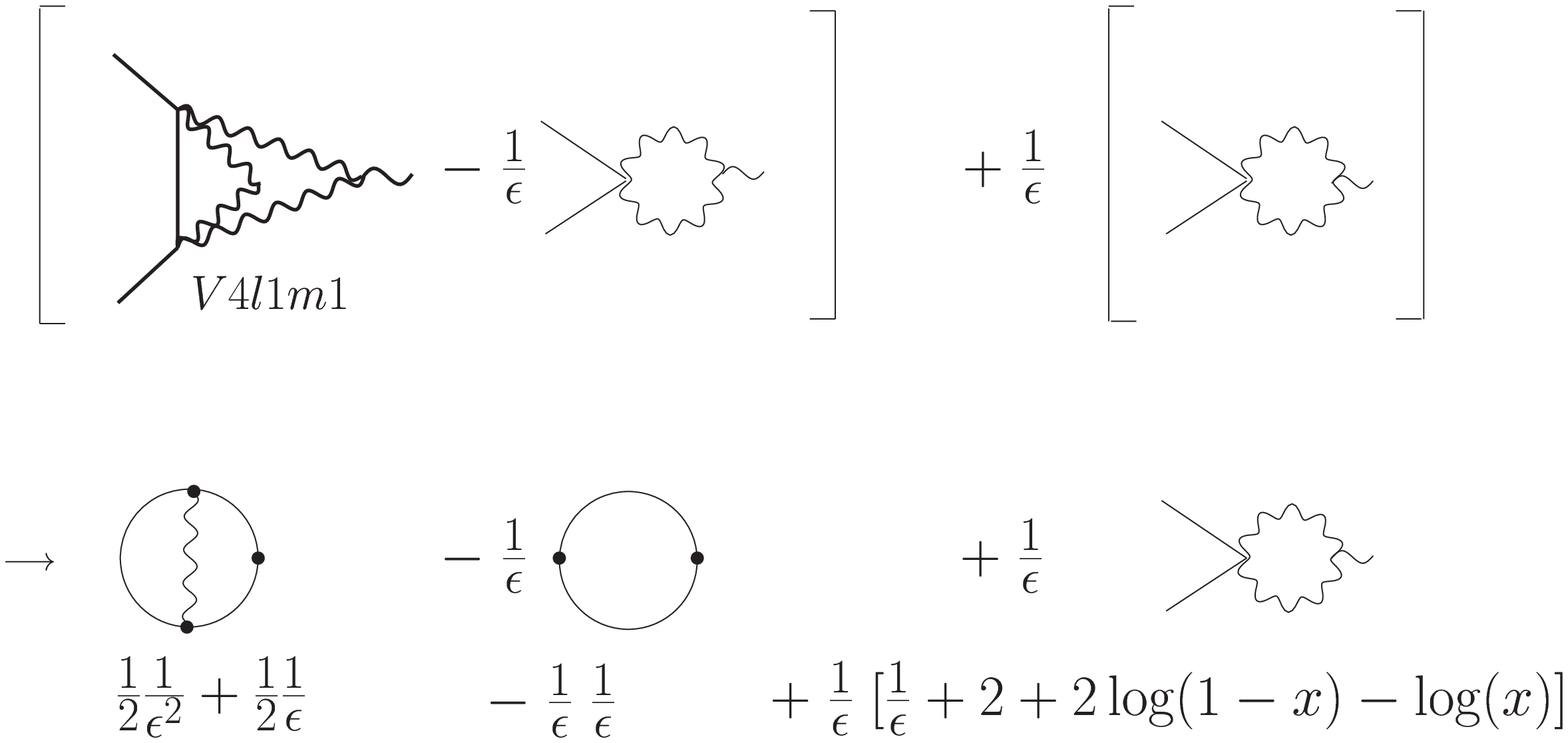,width=10cm,height=5cm}
     \caption{The divergences of the diagram V4l1m1 can be obtained by renormalization theory methods.
} 
     \label{c3}
     \end{figure}

The diagram V4l1m1, drawn in bold lines in Fig. \ref{c3},  is UV divergent.

\noindent
Let us introduce a subtraction as in Fig. \ref{c3}.
This subtraction is actually a counter term for the subdivergence in
the $\overline{MS}$ scheme.  It is known\cite{col} that after such a 
renormalization of subdivergences the divergences of the result
are polynomials in dimensionful parameters.
Since the diagram is dimensionless, they are just constants
and we can  replace massless lines by massive ones and set  the external
momentum to zero, in this way
avoiding spurious IR divergences.
This replacement is shown in the second row of Fig. \ref{c3} and
allows to calculate the singular part of the diagram.
For this, we also have to add the subtraction at arbitrary $s$,
$x = \frac{\sqrt{-s+4}-\sqrt{-s}}{\sqrt{-s+4}+\sqrt{-s}}$ (second square bracket in Fig.~\ref{c3}).
The final result agrees with\cite{ll04} and\cite{web-masters:2004nn} 
for the MI V4l1m1. 
\subsection{Expressing  a  subtracted subdiagram by a dispersion relation}  
\begin{figure}[ht] 
\epsfig{file=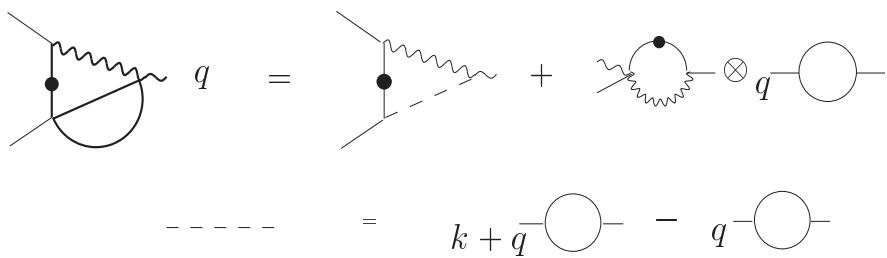,width=11.9cm,height=3.4cm}
     \caption{The IR divergent 2-loop diagram V4l3md with a subtraction.
} 
     \label{c4}
     \end{figure}

The diagram V4l3md contains an IR singular subloop 
when the dotted line becomes on-shell. 
After integrating out the UV divergent 2-point subloop, we get
 \begin{eqnarray}
I_{V4l3md} = \int \frac{d^dk}{[k-p_1^2-m^2]^2k^2} B_0[(k+q)^2,m^2,m^2].
 \end{eqnarray}
By subtracting and adding the 2-point subloop
as shown schematically in Fig. \ref {c4} we get a finite vertex type
integral, plus a product of 2-point functions.

\noindent
The vertex integral may be further rewritten using a dispersion relation
representation for the difference
$B_0[(k+q)^2,m^2,m^2] - B_0(q^2,m^2,m^2)$ under the integral.
After all these preparations, the $\epsilon$ expansion of the diagram
may be determined.

\vspace{.1cm}

Methods like the ones presented in this section  are also applicable for 4-point
functions.

\end{document}